\documentclass[intlimits,twoside,a4paper]{article}
 \usepackage[cp1251]{inputenc}
 
\usepackage[eqsecnum]{cmpj3}

 \usepackage{bm}
\usepackage{graphicx}
\issue{2020}{23}{2}{23608}
\doinumber{10.5488/CMP.23.23608}
 \title[On theory of magnetorheological effect in elastomers]{Magnetorheological effect in elastomers containing uniaxial ferromagnetic particles}

\author[V.M. Kalita, I.M. Ivanova, V.M. Loktev]{V.M. Kalita\refaddr{label1,label2, label3}, I.M. Ivanova\refaddr{label1}, V.M. Loktev\refaddr{label1,label4}}
\addresses{
\addr{label1} National Technical University of Ukraine ``Igor Sikorsky Kyiv Polytechnic Institute'',\\ 37 Peremohy Ave., 03056 Kyiv, Ukraine
\addr{label2} Institute of Physics of the National Academy of Sciences of Ukraine, 46 Nauky Ave., 03028 Kyiv, Ukraine
\addr{label3} Institute of Magnetism of the National Academy of Sciences of Ukraine and  Ministry of Education of Ukraine, 36-b Vernadsky Blvd., 03142 Kyiv, Ukraine
\addr{label4} Bogolyubov Institute for Theoretical Physics  of the National Academy of Sciences of Ukraine,\\  14-b Metrolohichna St.,  03143 Kyiv, Ukraine }

\date{Received January 30, 2020, in final form April 14, 2020}

\begin{document}

\maketitle

\begin{abstract}
The description of the collective magnetorheological effect induced by magnetic field in magnetoactive elastomers is proposed. The condition of consistency is used between magnetic and mechanic momenta of forces  exerted on magnetically uniaxial ferromagnetic particles in elastomer at their magnetization. The study shows that even in the case of small concentration of particles, the value of magnetically-induced shear can be anomalously large, reaching up to tens of percent. The deformation of magnetoactive elastomer can evolve critically, as a second-order phase transition, if magnetic field is aligned along the easy axis of particles.
\end{abstract}

  \section{Introduction}

Usually elastomers are a synthetic nonmagnetic (composite, high-molecular) materials that can easily and reversibly change their form or size under the external influence –- thermal, mechanical or field. If some elastomer contains magnetic, mainly ferromagnetic, particles, then it responds to the external magnetic field. Thus, it is referred to as magnetoactive elastomer (MAE)  \cite{Farshad,Nadzharyan1}. The magnetic field affects MAE through its magnetic subsystem, understandably \cite{Menzel, Odenbach, Cantera,Ivaneyko,Sanchez}. MAE can not only be deformed, even substantially, by tens of percent, but also can change its physical properties and characteristics \cite{Stepanov1, Belyaeva, Bodnaruk1, Stepanov2,Bodnaruk2, Stoll}. The change of MAE elastic modules has been defined as the magnetorheological effect (MRE) \cite{Lokander, Varga, Nadzharyan2, Yao, Watanabe}. Due to all their abovementioned properties, MAE are now classified as smart materials  \cite{Sutrisno,Sanchez2}.

The MRE caused by the rotation of magnetically uniaxial ferromagnetic particles  in MAE was examined in reference \cite{Kalita1}. However, in this paper the particles were considered of discus form for calculation simplification, although the spherical form of particles is most common in usual experimental investigations of MAE (see references \cite{Farshad, Nadzharyan1,Menzel, Odenbach}).

Both  soft and hard ferromangentic particles of different forms with different matrixes can be used as an inclusion in MAE \cite{Lokander,Sternberg}. However, magneto-rheological effect in MAE with soft magnetic particles has larger reversibility at magnetization.  The best way to observe an MRE is in matrixes formed with elastomers with the lower value of elastic modules \cite{Belyaeva}. The spherical particles of magnetically soft carbonyl iron are most commonly used as a filler in MAE, although MRE can be also observed for nonspherical particles \cite{Sternberg}. The sizes of filler magnetic particles can be different and have bimodal distribution \cite{Lokander,Sternberg,Sorokin}.

The purpose of our paper is to examine MAE in the low range of concentration of particles. In this case, one can assume that the rotation of particles has a linear input in MRE, while all other mechanisms (i.e., magneto-dipole interaction of particles) have at least quadratic dependence on concentration. Thus, MAE with small concentration of particles can be used as objects to experimentally prove and, furthermore, to measure how field-induced rotation of ferromagnetic particles can influence the value of MAE shear modulus, and in this way to describe MRE quantitively.

The largest value of MRE can be observed at concentration of particles in filler at around 30\%.~
\cite{Belyaeva,Lokander,Sternberg,Sorokin} and the main interactions are dipole magnetic pair interactions, under the influence of which particles shift and form chains \cite{Menzel,Odenbach}. However, even in the case of large concentrations of particles in MAE, as stated in \cite{Sternberg}, the magnetic anisotropy should be taken into account \cite{Kalita4}. It is proposed in~\cite{Bodnaruk3}, that at high concentrations particles would not only shift but also rotate if they have a magnetic anisotropy. For MAE with low concentration of magnetically anisotropic particles, the MRE is mainly influenced by the rotation of particles induced by the magnetic field.

We will also show that due to the  critical influence of magnetic field on the shear absolute value, its change has some indications of a phase transition. The effect of critical rotation of particles was previously observed on the example of a single ferromagnetic particle in elastomer \cite{Kalita2}, although collective particles dynamics and its consequences are more relevant with regard to the real systems.

This paper is executed and dedicated to the 60-year jubilee of the well-known Ukrainian theoretical physicist, professor Ihor Mryglod --- specialist in physics of phase transitions and soft matter theory, to which the most elastomer compounds can be related.

  \section{Free energy and equation of state}

\begin{figure}[!h]
	\centerline{\includegraphics[width=0.6 \textwidth]{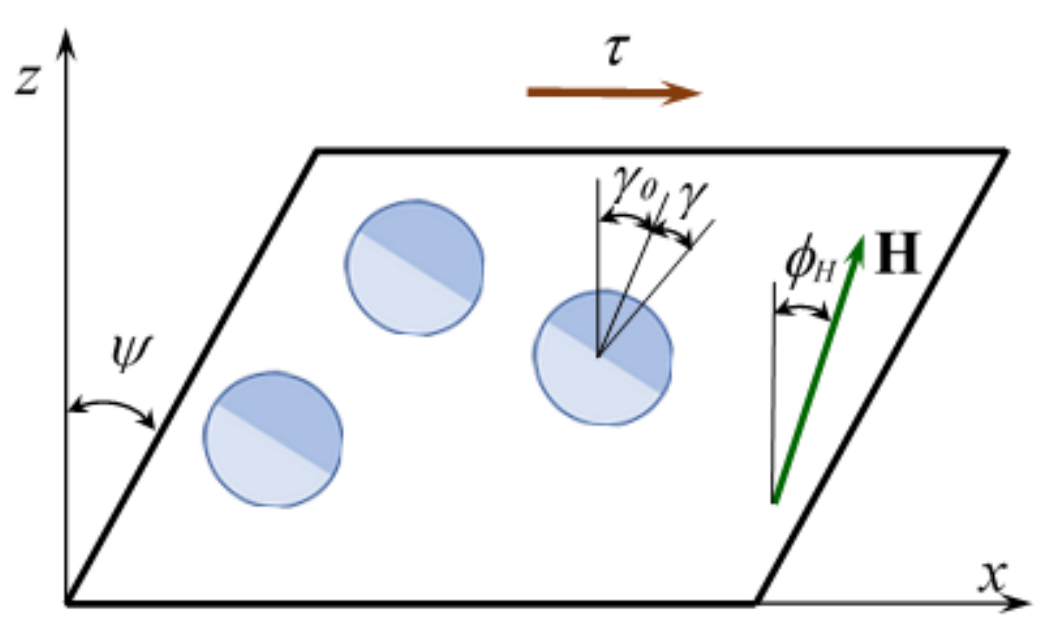}}
	\caption{(Colour online) MAE with spherical particles at $\varphi_H$ angle with respect to $\textbf{H}$ field. 
	The particles easy axes were directed at $\gamma_0$ angle before magnetization, after the magnetization they are directed at $\gamma$ angle.
	The shear value is $\psi$, the tangent stress is $\tau$. The initial coordinate system exists, with the axis $x$ defined by $\tau$ and the axis $z$ is perpendicular to it, as proposed in the text.}
	\label{Fig1}
\end{figure}

Let us observe a model of MAE, where magnetically uniaxial (or simply, easy-axial) ferromagnetic spherical particles with parallel magnetic anisotropy axes\footnote{The parallel fixation of these axes is easily achievable, in fact, while obtaining elastomers from their liquid phase in external magnetic field. The resulting condensed elastomer, which can usually  exist in a wide range of temperatures, has particles randomly placed in its volume  with rigidly fixed and collinear magnetic anisotropy axes.} are dispersed randomly, as shown in figure~\ref{Fig1}, and are fixed in a matrix. Particles as per usual are considered to be polydomain (i.e., with magnetic momentum equal to zero), and their anisotropy easy axes are directed at $\gamma_0$ angle. The particle turns by $\gamma$ angle after its magnetization in the magnetic field, directed at  $\varphi_H$ angle, and shear strain $\psi$, caused by tangent stress (all angles are calculated at the assumption of existence of a natural coordinate system;  for example, \emph{x} axis is determined by the applied stress $\tau$, and $z$ axis is perpendicular to the surface).

Further, let us assume that a magnetic field \textbf{H} of random direction and a tangent stress $\tau$  are applied to the sample. On the one hand, the particle under the field and stress influence forms its magnetization direction, and on the other hand, simultaneously shears and rotates at $\gamma$ angle from its initial position in the elastomer.

The magnetic energy density of MAE unit volume at low  concentration of  particles is equal to the product of single particles energy \cite{Kalita2} and the number of particles in the volume
\begin{equation}
\label{Eq1}
E_{\textrm{mag}}=-n\left[ \frac{1}{2}\chi_{\parallel}H^{2}\cos^2{\left(\gamma_0+\gamma-\varphi_{H}\right)}+
\frac{1}{2}\chi_{\perp}H^{2}\sin^2{\left(\gamma_0+\gamma-\varphi_{H}\right)}
 \right]V_0\,,
  \end{equation}
  where $n$ is the number of particles per unit volume (concentration), and $n \ll 1$ that permits to neglect the mutual effect of particles due to their magnetization, $V_0$ is the volume of a particle, $\chi_{\parallel}$ is the component of the magnetic susceptibility along the easy axis of magnetization, $\chi_{\perp}$ is the component of the magnetic susceptibility perpendicular to the easy axis, $\chi_{\parallel}>\chi_{\perp}$.

 If the particles are of nonspherical form, for example they are rotation ellipsoids, then the rotation axis of this ellipsoid can be an easy-magnetization axis.

Relation (\ref{Eq1}) fulfills in the magnetic field that is lower than the maximum demagnetizing field of the particle   $H<H_{\textrm{dem}}^{\textrm{max}} =Nm_{\textrm{S}}$, where $m_{\textrm{S}}$ is the magnetization saturation of the particle,  $N$ is its demagnetization factor that for spherical particles is equal to  $N=4/(3\piup)$,  $\chi_{\parallel}^{-1}=N$  and $\chi_{\perp}^{-1}=N+H_{\textrm{A}}/m_{\textrm{S}}$;  $H_{\textrm{A}}$  is the magnetic anisotropy field.

The elastic energy of elastomer with the rotating spherical particle was calculated in linear approximation~\cite{Phan,Puljiz} in previous papers \cite{Kalita2,Ohayon}. So, the density of elastic energy can be considered as:
\begin{equation}
\label{Eq2}
E_{\textrm{el}}=n4\piup\mu R^3 \left(\frac{\psi}{2}-\gamma\right)^2,
\end{equation}
where $\mu$ is the shear modulus for MAE without inclusions.

The quadratic angle dependence of $E_\text{el}$ retains for particles of nonspherical form while the value of coefficient of proportionality rises.

It is taken into account in equation (\ref{Eq2}) that the sample as a whole uniformly rotates by $\psi/2$ angle with shear, as shown in figure \ref{Fig1} \cite{Kalita1}. The relative volume concentration of particles is equal to $p=nV_0$. The density of particles total energy of magnetized MAE under mechanical shear strain can be written as
\begin{eqnarray}
\label{Eq3}
E_{\text{sum}}=-\frac{1}{2}pH^{2}\left[\chi_{\perp}+\Delta\chi\cos^2{(\gamma_0+\gamma-\varphi_H)}\right] +3p\mu\left(\frac{\psi}{2}-\gamma\right)^{2}
+\frac{1}{2} G^{\textrm{eff}} \psi^2-\tau\psi,
\end{eqnarray}
where $G^{\textrm{eff}}=G^{\textrm{eff}} (H=0)$ is the effective shear modulus for MAE in the absence of field, $H=0$ and $\Delta\chi=\chi_{\parallel}-\chi_{\perp}$.

The equation of state can take the following form:
\begin{equation}
\label{Eq4}
\frac{\partial E_{\text{sum}}}{\partial\gamma}=\frac{1}{2}pH^2\Delta\chi\sin2{(\gamma_0+\gamma-\varphi_H)}-6p\mu\left(\frac{\psi}{2}-\gamma\right)=0,
\end{equation}
\begin{equation}
\label{Eq5}
\frac{\partial E_{\textrm{sum}}}{\partial\psi}=3p\mu\left(\frac{\psi}{2}-\gamma\right)+ G^{\textrm{eff}} \psi-\tau=0.
\end{equation}

The torque generated by magnetic field  acting upon a particle  is compensated in  equation (\ref{Eq4}) by the torque that is coming from the matrix.

Thus, it follows from equation (\ref{Eq5}) that
\begin{equation}
\label{Eq6}
\psi=\frac{\tau+3p\mu\gamma}{G^{\textrm{eff}}+\frac{3}{2}p\mu}\,.
\end{equation}

The density of total energy (\ref{Eq3}) can be written, using equation (\ref{Eq6}), as a function of the single variable $\gamma$:
\begin{equation}
\label{Eq7}	
E_{\textrm{sum}}=-\frac{1}{2}pH^2\left[\chi_{\perp}+\Delta\chi\cos^2{(\gamma_0+\gamma-\varphi_H)}\right]+\frac{6p\mu G^{\textrm{eff}}\gamma^2-6p\mu\gamma\tau-\tau^2}{2(G^{\textrm{eff}}+\frac{3}{2}p\mu)}\,.
\end{equation}	

In such a case, the equation of state, derived from substitution of the equation (\ref{Eq6}) into the equation~(\ref{Eq4}), takes the form
\begin{equation}
\label{Eq8}	
\frac{1}{2}\Delta\chi H^2\sin{2(\gamma_0+\gamma-\varphi_H)}+6\frac{\mu G^{\textrm{eff}}}{G^{\textrm{eff}}+\frac{3}{2}p\mu}\gamma=3\frac{\mu}{G^{\textrm{eff}}+\frac{3}{2}p\mu}\tau\,.
\end{equation}	

When $\varphi_H=0$, $\tau=0$, equation (\ref{Eq8}) transforms into
\begin{equation}
\label{Eq9}	
\frac{H^2}{H^2_\text{A}}\sin{2(\gamma_0+\gamma)}+12\frac{\mu}{\Delta\chi H^2_\text{A}}\frac{G^{\textrm{eff}}/\mu}{G^{\textrm{eff}}/\mu+3/2p}\gamma=0.
\end{equation}	

\begin{figure}[!b]
\centerline{\includegraphics[width=0.6 \textwidth]{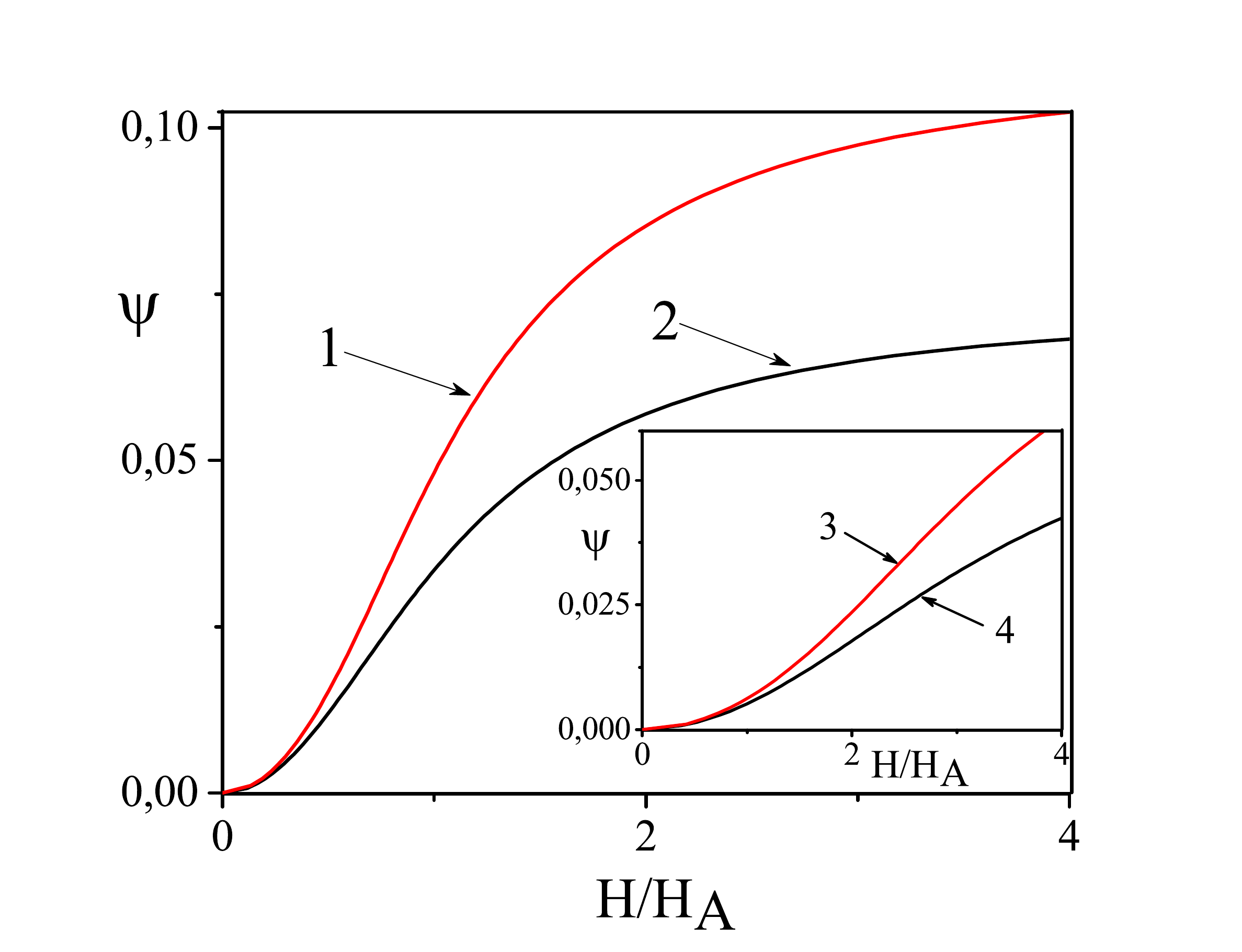}}
\caption{(Colour online) The shear $\psi$ vs normed magnetic field $H/H_{\textrm{A}}$. Curves 1 and 2 are plotted for  $\varphi_H=\piup/4, \piup/6$ and $\mu/\Delta\chi H_{\textrm{A}}^2=0.2.$ The same dependencies at a harder matrix $\mu/\Delta \chi H_{\textrm{A}}^2=2$ are on the insert.}
\label{Fig2}
     \end{figure}

The obtained equation (\ref{Eq9}) allows one to determine the rotation angle of particles at their magnetization. Despite the absence of the tangent stress $\tau=0$,  the rotation of particles also occurs in the magnetic field, because it is caused by the change of magnetic anisotropy axes direction, that, in turn, causes the shear deformation
\begin{equation}
\label{Eq10}
\psi=\frac{3p}{G^{\textrm{eff}}/\mu+\frac{3}{2}p}\gamma	
\end{equation}
of MAE as an elastic medium.

 Figure \ref{Fig2}  shows the dependencies of magnetically-induced shear in MAE with   $p=0.05$, obtained from quantitative solutions of (\ref{Eq4}), (\ref{Eq5})  in the absence of external stress $\tau=0$ at $\gamma_0=0$ for two magnetic field orientations: $\varphi_H=\piup/6$  and $\varphi_H=\piup/4$. The graph is executed for fields, normalized on $H_{\textrm{A}}$, that are lower than maximal particles demagnetization field.  As is in reference \cite{Kalita3}, here the magnetization was assumed equal to $m_{\textrm{S}}\sim 10^3$ emu/cm$^3$,  the anisotropy  field is $H_{\textrm{A}} \sim 1$ kOe. Thus, the difference between reciprocal susceptibilities is equal to  $\chi^{-1}_{\parallel}-\chi^{-1}_{\perp}=H_{\textrm{A}}/m_{\textrm{S}}\approx 1 $ Oe$\, \cdot\, $cm$^3$/emu. For soft elastic matrix with modulus $\mu=1 \cdot 10^3$ J/m$^3$, the ratio between elastic modulus and anisotropy energy is much less than one $\mu/\Delta\chi H_{\textrm{A}}^2=0.2$. The curves on the insert are plotted for elastic hard matrix  $\mu=1\cdot10^4 $~J/m$^3$.

This figure also shows  that magnetically induced shear reaches its saturation in elastic soft matrix.  In elastic hard matrix, the shear evolves slower with the field growth (see insert on figure \ref{Fig2}) and  the rotation of particles does not reach its maximum at $H<H_{\textrm{dem}}^{\textrm{max}}$\,.

Thus, the field-induced striction in MAE with particles that have identically oriented axes of magnetic anisotropy can be abnormally large, reaching up to tens of percent.

\section{Magnetic field influence on shear absolute value}

It is interesting to consider the  equation (\ref{Eq8}) for lower fields, which linearizes in the range of  $H\rightarrow0$, where $\gamma\rightarrow0$, and takes the following form:
\begin{eqnarray}
\label{Eq11}
\frac{1}{2}p\Delta\chi H^2\left[\sin{2(\gamma_0-\varphi_H)}+2\gamma\cos{2(\gamma_0-\varphi_H)}\right]
+6p\frac{\mu G^{\textrm{eff}}}{G^{\textrm{eff}}+\frac{3}{2}p\mu}\gamma
=3\frac{p\mu}{G^{\text{eff}}+\frac{3}{2}p\mu}\tau.
\end{eqnarray}

The solution of equation (\ref{Eq11}) yields the expression for the rotation angle of particles in MAE
\begin{equation}
\label{Eq12}
\gamma=\frac{3\frac{\mu}{G^{\textrm{eff}}+\frac{3}{2}p\mu}\tau-\frac{1}{2}\Delta\chi H^2\sin{2(\gamma_0-\varphi_H)}}{6\frac{\mu G^{\textrm{eff}}}{G^{\textrm{eff}}+\frac{3}{2}p\mu}+\Delta\chi H^2\cos{2(\gamma_0-\varphi_H)}}\,.
\end{equation}
	
Such a rotation can be followed by the inevitable shear in MAE
\begin{equation}
\label{Eq13}
\psi=\frac{1}{G^{\textrm{eff}}+\frac{3}{2}p\mu}\left[\tau+3p\mu\frac{3\mu\tau-\frac{1}{2}(G^{\textrm{eff}}+\frac{3}{2}p\mu)\Delta\chi H^2\sin{2(\gamma_0-\varphi_H)}}
{6\mu G^{\textrm{eff}}+(G^{\textrm{eff}}+\frac{3}{2}p\mu)\Delta\chi H^2\cos{2(\gamma_0-\varphi_H)}}\right].
\end{equation}

The second term in the brackets of equation (\ref{Eq13}) describes the magnetic fields influence on the shear in MAE. For example, in the low fields $H$, one obtains
\begin{equation}
\label{Eq14}
\psi=\frac{\tau}{G^{\textrm{eff}}}\left[1-\frac{p\Delta\chi H^2\cos{2(\gamma_0-\varphi_H)}}{4G^{\textrm{eff}}}\right]-\frac{p\Delta\chi H^2\sin{2(\gamma_0-\varphi_H)}}{4G^\text{eff}}.
\end{equation}

This equation demonstrates that at all field \textbf{H} directions, tilted towards easy-magnetization axis  $\gamma_0\neq\varphi_H$, this field induces shear in MAE, except for the case of \textbf{H} perpendicular to particles easy axis,  or $\gamma_0-\varphi_H=\piup/2$\,.

The first term in the equation (\ref{Eq14}) indicates that the value of effective  modulus of shear is proportional to the magnetic field strength squared in the lower magnetic fields range:
\begin{equation}
\label{Eq15}
G^{\textrm{eff}}(H)=G^{\textrm{eff}}\left[1+\frac{p\Delta\chi H^2\cos{2(\gamma_0-\varphi_H)}}{4G^{\textrm{eff}}}\right].
\end{equation}

If the field is directed along particles easy-magnetization axis, then it enlarges the value of effective modulus of shear, at the same time counteracting the shear. And vice versa: if this field is perpendicular to the easy axis, then at $\tau\neq0$, the value of effective modulus of shear decreases during magnetization.

It is interesting to note that the field component in the effective modulus of shear is defined by the ratio between MAE matter constants, $(\Delta\chi H_{\textrm{A}}^2)/G^{\textrm{eff}}$, and its value can exceed 1. The largest growth of the modulus of shear will be in the magnetic field if, for example, $\gamma_0=\varphi_H=0$.

Next, the maximal growth of shear modulus is defined for all values of the magnetic field. One can see that the equation of state (\ref{Eq8}) at $\gamma_0=\varphi_H=0$ comes down to
\begin{equation}
\label{Eq16}
\frac{1}{2}\Delta\chi H^2\sin{2\gamma}+6\frac{\mu G^{\textrm{eff}}}{G^{\textrm{eff}}+\frac{3}{2}p\mu}\gamma=3\frac{\mu}{G^{\textrm{eff}}+\frac{3}{2}p\mu}\tau.
\end{equation}

If the tangent stress is small, $\tau\rightarrow0$,  then rotation of particles caused by it is also small, $\gamma \rightarrow 0$. This equation of state shows that in the above-mentioned case, the value of particles rotation angle is directly proportional to the external stress and inversely proportional to the field strength squared:
\begin{equation}
\label{Eq17}
\gamma=\frac{3\mu\tau}{6\mu G^{\textrm{eff}}+\Delta\chi H^2(G^{\textrm{eff}}+\frac{3}{2}p\mu)}\,.
\end{equation}

If $\gamma_0=\varphi_H=0$ condition is satisfied, then the value of shear is defined as
\begin{equation}
\label{Eq18}
\psi=\frac{\tau}{G^\text{eff}+\frac{3}{2}p\mu}\left[1+3p\mu\frac{3\mu}{6\mu G^{\textrm{eff}}+\Delta\chi H^2(G^{\textrm{eff}}+\frac{3}{2}p\mu)}\right].
\end{equation}

Thus, the field dependence of effective modulus of shear  can be written as follows, in case $\tau \rightarrow0$:
\begin{equation}
\label{Eq19}
G^{\textrm{eff}}(H)=G^{\textrm{eff}}\frac{6\mu +\Delta\chi H^2(1+\frac{3}{2G^{\textrm{eff}}}p\mu)}{6\mu +\Delta\chi H^2}.
\end{equation}

It is easy to see here that modulus of shear is proportional to the magnetic field strength squared in the lower field range, while in the higher field this dependence comes to its saturation. As it follows from equation (\ref{Eq19}), in the external magnetic field, the MAE effective shear  modulus (under strong inequality $p\ll1$)  reaches its maximum value
\begin{equation}
\label{Eq20}
G^{\textrm{eff}}_{\textrm{max}}=G^{\textrm{eff}}\left(1+\frac{3p\mu}{2G^{\textrm{eff}}}\right).
\end{equation}

Thus, we have obtained, as it is seen from equation (\ref{Eq20}), that magnetically induced change of shear modulus, caused by rotation of magnetic particles in elastomer, is directly proportional to the  concentration of particles $p$.

\section{Critical behaviour}

Let us assume that $\gamma_0=0$, $\varphi_H=\piup/2$. Such directions correspond to the case where easy axes of particles  are perpendicular to the magnetic field, the field itself lies on the surface of MAE and is parallel to the tangent stress (see figure~\ref{Fig1}). This geometry leads, from the equation (\ref{Eq8}), to the equation of state of the following form:
\begin{equation}
\label{Eq21}
-\frac{1}{2}\Delta\chi H^2\sin{2\gamma}+6\frac{\mu G^{\textrm{eff}}}{G^{\textrm{eff}}+\frac{3}{2}p\mu}\gamma=3\frac{\mu}{G^{\textrm{eff}}+\frac{3}{2}p\mu}\tau.
\end{equation}

The foregoing equation, presented in polynomial form
\begin{equation}
\label{Eq22}
\left(6\frac{\mu G^{\textrm{eff}}}{G^{\textrm{eff}}+\frac{3}{2}p\mu}-\Delta\chi H^2\right)\gamma+\frac{2}{3}\Delta\chi H^2\gamma^3=3\frac{\mu}{G^{\textrm{eff}}+\frac{3}{2}p\mu}\tau,
\end{equation}
shows the presence of some critical field
\begin{equation}
\label{Eq23}
H_{\textrm{cr}}=\sqrt{6\frac{\mu G^{\textrm{eff}}}{\Delta\chi(G^{\textrm{eff}}+\frac{3}{2}p\mu)}}.
\end{equation}

In the case of concentration of particles  approaching zero, $p\rightarrow0$, the value of critical field in MAE is equal, as it should be, to the value of critical field for one particle \cite{Kalita2}.

 Thus, the rotation angle of particles has different field dependencies in the different range of field, if $\tau=0$:
\begin{equation}
\label{Eq24}
\gamma=0, \quad
 \text{if} \quad H<H_{\textrm{cr}}\,,
\end{equation}	
\begin{equation}
\label{Eq25}
\gamma=\pm\sqrt{3\frac{H-H_{\textrm{cr}}}{H_{\textrm{cr}}}}\,, \quad \text{if}\quad H>H_{\textrm{cr}}\,.
\end{equation}

\begin{figure}[!t]
\centerline{\includegraphics[width=0.58 \textwidth]{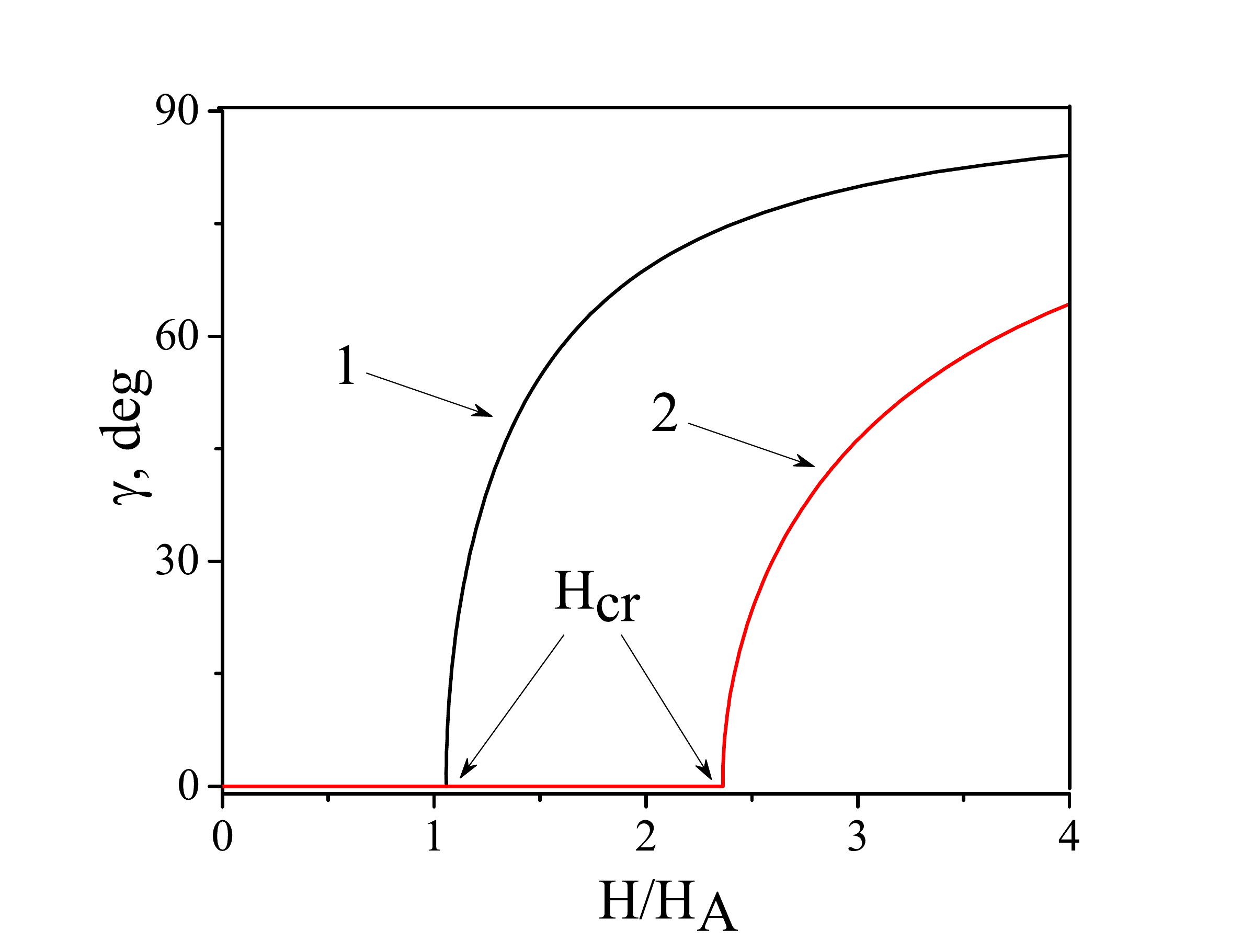}}
\caption{(Colour online) Angle $\gamma$ of rotation of a particle vs field for $p=0.05$, $\gamma_0=0$,  $\varphi_H=\piup/2$ and $\tau=0$. Curves 1 and 2 are plotted for $\mu/\Delta\chi H_{\textrm{A}}^2=0.2,1$.}
\label{Fig3}
    \end{figure}

Figure \ref{Fig3} shows dependencies of $\gamma$ rotation angle of particles in MAE with  $p=0.05$, derived from the solution of  the equation of state (\ref{Eq9}), for $\gamma_0=0$,  $\varphi_H=\piup/2$ and $\tau=0$. The curve 1 corresponds to the solution obtained at $\mu/\Delta\chi H_{\textrm{A}}^2=0.2$, and curve 2 is obtained for $\mu/\Delta\chi H_{\textrm{A}}^2=1$ ratio.

According to equation (\ref{Eq25}), particles can rotate both clockwise and counterclockwise. That is why we can assert that for the case of $\gamma_0=0$,  $\varphi_H=\piup/2$,  the  rotation of particles leads to a jump-like change of magnetic susceptibility value. The change is not followed by the shear of sample, because particles can have both positive and negative angles of rotation with an equal probability.

Small tangent stress should be introduced for the observation of critical shear deformation in MAE, originated by critical rotation of particles, to exclude uncertainty in the direction of particle rotations. Thus, the rotation of particles in the range of field $H<H_{\textrm{cr}}$  can be defined as
\begin{equation}
\label{Eq26}
\gamma(H<H_{\textrm{cr}})=\frac{3^2}{2\phantom{!}}\frac{\mu\tau}{\left(G^{\textrm{eff}}+\frac{3}{2}p\mu\right)\Delta\chi H_{\textrm{cr}}(H_{\textrm{cr}}-H)}\,.
\end{equation}	

The expression for shear in the same range of field $H<H_{\textrm{cr}}$ is:
\begin{equation}
\label{Eq27}
\psi(H<H_{\textrm{cr}})=\frac{\tau}{G^{\textrm{eff}}+\frac{3}{2}p\mu}\left[1+\frac{3^3}{2\phantom{!}}\frac{p\mu^2}{\left(G^{\textrm{eff}}+\frac{3}{2}p\mu\right)\Delta\chi H_{\textrm{cr}}(H_{\textrm{cr}}-H)}\right].
\end{equation}

The derivative, $\rd\psi/\rd\tau$, also has a critical field dependence at $H\rightarrow H_{\textrm{cr}}$, as it follows from the second term of equation (\ref{Eq27}):
\begin{equation}
\label{Eq28}
\frac{\rd\psi}{\rd\tau}\bigg|_{H\rightarrow H_{\textrm{cr}}}\sim \frac{1}{H_{\textrm{cr}}-H}\,.
\end{equation}

Thus, this derivative has a peculiarity in critical point at lower field range and the value of effective modulus of shear decreases near the critical point.

\section{Conclusions}

 The peculiarities in the behaviour of MAE in the case of small concentration of ferromagnetic particles with easy-axis magnetic anisotropy are studied above, using approximation of linear elasticity problem. The obtained results correctly describe, at a qualitative level, the field dependencies of the shear and the effective modulus of shear, which are proportional to the magnetic field strength squared in the lower magnetic fields range, $H\rightarrow0$,  and come to saturation in higher fields.

If the magnetic field is aligned along the anisotropy axis, then it leads to an increase of the MAE effective modulus of shear, in particular. While the magnetic field is directed perpendicularly to magnetization easy axis of particles and is low, the derivative of shear with respect to tangent stress has a peculiarity near the critical point, where MAE undergoes a magneto-elastic transition.  The spontaneous shear in MAE occurs in the critical point itself, and has all features of a second-order phase transition. Its description appears to be similar to the one in Landau phase transition theory, where everything is defined by the equations of state. The peculiarity in MAE is based on the fact that magnetization easy axes of particles start to rotate in the critical point. This rotation causes the rotation of particles and, vice versa, the rotation of particles induces the magnetization rotation. The magnetization starts to tilt towards easy axis, while the system approaches its critical point. There is no doubt that the obtained results are not difficult to verify experimentally in order to clarify the  range of parameters, where the supposition about the independence of elastomer particles  is valid.

\section{Acknowledgements}

The work by V.M.~Loktev was partially supported by grants No. 0117U000236 and
No. 0117U000240 from the Department of Physics and Astronomy of the
National Academy of Sciences of Ukraine, and the collaboration under
the Ukrainian-Israeli Scientific Research Program of the Ministry of
Education and Science of Ukraine (MESU) and the Ministry of Science and
Technology of the State of Israel (MOST).

\ukrainianpart

\title{Магніто-реологічний ефект у еластомерах з одновісними феромагнітними частинками}
\author{В.М. Каліта\refaddr{label1,label2,label3}, І.М. Іванова\refaddr{label1}, В.М. Локтєв\refaddr{label1,label4}}
\addresses{
	\addr{label1} Національний технічний університет України ``Київський  політехнічний інститут імені Ігоря Сікорського'', просп. Перемоги, 37, 03056 Київ, Україна
	\addr{label2} Інститут фізики НАН України, просп. Науки, 46, 03028 Київ, Україна
	\addr{label3} Інститут магнетизму НАН України та МОН України, бул. акад. Вернадського, 36-б, 03142 Київ, Україна
	\addr{label4} Інститут теоретичної фізики імені М.М. Боголюбова НАН України,\\ вул. Метрологічна, 14-б,  03143 Київ, Україна}
\makeukrtitle

\begin{abstract}
	\tolerance=3000%
	Теоретично описано явище магніто-реологічного ефекту в магнітоактивних еластомерах, що індукується магнітним полем. Використана умова узгодженості  механічного і магнітного моментів сил, що діють на одновісні феромагнітні частинки в магнітоактивному еластомері при їх намагнічуванні. Отримано, що у випадку малих концентрацій частинок, величина вимушеного зсуву, що індукується магнітним полем, може бути аномально великою і сягати десятків відсотків. У випадку, якщо магнітне поле напрямлене перпендикулярно по відношенню до легких вісей частинок, деформація магнітоактивного еластомеру може розвиватися критично, подібно до фазового переходу другого роду.
	\keywords магнітоактивні еластомери, магніто-реологічний ефект, магнітострикція, критичний зсув
	
\end{abstract}

      \end{document}